\documentclass[12pt,preprint]{aastex}
\usepackage{emulateapj5}
\usepackage{apjfonts}

\newcommand{\HI}{H\,{\sc i}} 

\newcommand{\kms}{${\rm km~s^{-1}}$}

\slugcomment{ }

\shortauthors{McCLURE-GRIFFITHS \& GAENSLER}
\shorttitle{DISTANCE TO SGR~1806--20}

\begin{document} 

\title{Constraints on the Distance to SGR~1806--20 from \HI\ Absorption}

\author{N.\ M.\ McClure-Griffiths\altaffilmark{1} and B.\ M.\ Gaensler\altaffilmark{2}}
\altaffiltext{1}{Australia Telescope National Facility, CSIRO, PO Box 76, Epping NSW 1710, Australia; naomi.mcclure-griffiths@csiro.au} 
\altaffiltext{2}{Harvard-Smithsonian Center for Astrophysics, 60 Garden
Street, Cambridge, MA 02138, USA; bgaensler@cfa.harvard.edu}

\authoraddr{Address correspondence regarding this manuscript to: 
                N. M. McClure-Griffiths
                ATNF, CSIRO
                PO Box 76
		Epping NSW 1710
		Australia }

\begin{abstract}

The giant flare detected from the magnetar SGR~1806--20 on 2004
December 27 had a fluence more than 100 times higher than the only two
other SGR flares ever recorded.  Whereas the fluence is independent
of distance,
an estimate for the luminosity of the burst depends on
the source's distance, which has previously been argued to be
$\sim15$~kpc.  The burst produced a bright radio afterglow, against
which \citet{cameron05} have measured an \HI\ absorption
spectrum. This has been used to propose a revised distance to
SGR~1806--20 of between 6.4 and 9.8~kpc.  Here we analyze this
absorption spectrum, and compare it both to \HI\ emission data from
the Southern Galactic Plane Survey and to archival $^{12}$CO survey
data.  We confirm $\sim$6~kpc, as a likely lower limit on the distance
to SGR~1806--20, but argue that it is difficult to place an upper
limit on the distance to SGR~1806--20 from the \HI\ data currently
available. The previous value of $\sim15$~kpc thus remains
the best estimate of the distance to the source.

\end{abstract}

\section{Introduction}
\label{sec:intro}

Magnetars are highly magnetized neutron stars, with inferred surface
magnetic fields $\ga10^{15}$~G \citep{dt92a,wt04}. Very rarely (perhaps
once per century per magnetar), a magnetar emits a giant flare of high
energy radiation, with luminosity $\ga10^{44}$~ergs.

On 2004 December 27, a brief but enormously bright burst of X-rays
and $\gamma$-rays was detected from the magnetar SGR~1806--20
\citep[e.g.,][]{bgm+04,pbg+05,hbs+05}. Spectrally and temporally, this giant
flare was very similar to the only two giant flares seen previously
(from SGR~0526--66 in 1979, and from SGR~1900+14 in 1998). However,  the flare
from SGR~1806--20 was much brighter than these
two prior events, with a fluence higher than from
any extra-solar event in recorded history. It was so intense that it
saturated all but a few of the instruments which detected it, and even
produced an observable echo off the moon \citep{mca+05}.

While the flux seen at Earth from this event was impressive, what
is more relevant for our overall understanding is the total luminosity
inferred for the flare, which depends in turn on the distance adopted
for the source.  Specifically, at a nominal distance to SGR~1806--20
of 15~kpc, the isotropic luminosity of the flare was
$\sim(2-4)\times10^{46}$~ergs \citep{pbg+05,hbs+05}. This is a
significant fraction of the total magnetic energy of the neutron
star, and suggests that we witnessed a catastrophic reconnection
event that may have reconfigured the magnetic field of the entire
star, and which perhaps only occurs a few times in a magnetar's
entire lifetime. Such a flare could have been detected by existing
instruments out to 30--40~Mpc \citep{pbg+05,hbs+05}.  This raises
the possibility that a small but significant fraction of the enigmatic
population of short gamma-ray bursts (GRBs) may be flares from
extragalactic magnetars \citep{ngpf05,hbs+05,pbg+05}.  On the other
hand, if the magnetar is at a smaller distance from us, then the
inferred energetics are substantially reduced, and flaring magnetars
are unlikely to be detectable in distant galaxies.

There have been several previous attempts to estimate the distance
to SGR~1806--20. \citet{cwd+97} directly estimated the distance to
this source via $^{12}$CO and X-ray observations.  They determined
the distribution of molecular clouds and their visual extinction
along this line of sight, and compared this to the foreground
hydrogen absorbing column inferred from the X-ray spectrum of
SGR~1806--20. By requiring the
hydrogen column inferred from visual extinction to match that seen
in X-ray absorption, \citet{cwd+97} concluded that the distance to
SGR~1806--20 is $14.5\pm1.4$~kpc.

Alternative distance estimates come more indirectly, by considering
the distance to the luminous blue variable (LBV)~1806--20 \citep[just
$12''$ to the east of SGR~1806--20;][]{hkc+99,kfk+02} and its
associated cluster of very massive stars. Detailed investigations
of the distance to LBV~1806--20 have been undertaken by \citet{ce04}
and \citet{fnk04}, who inferred distances to this star of
$15.1^{+1.8}_{-1.3}$~kpc and $\sim12$~kpc, respectively.  If magnetars
are associated with massive star formation
\citep[e.g.,][]{vhl+00,khg+04,gaensler05a}, then these distance
estimates can also potentially be extended to SGR~1806--20.

A new distance estimate has been made in the aftermath of the giant
flare from SGR~1806--20. The flare produced a rapidly fading radio
afterglow \citep{ck05,gaensler05b}, which for the first few days
after detection was sufficiently bright to be studied through \HI\
absorption.  Such a measurement was made by \cite{cameron05}, who
used the resulting absorption spectrum to argue for a hard lower
limit on the distance to SGR~1806--20 of 6.4~kpc, and to propose
an upper limit of 9.8~kpc.  The implied range is substantially
smaller than all previous estimates to this source, with immediate
implications for the energetics of the flare, for associations of
magnetars with massive stars, and for the nature of short GRBs.

Here we consider the distance to SGR~1806--20, using the absorption
data presented by \cite{cameron05}, $^{12}$CO data from the surveys
of \cite{sanders86} and \cite{dame01}, and additional information
on \HI\ emission in the field from the Southern Galactic Plane
Survey \citep[SGPS;][]{mcgriff05}.


\section{Results and Discussion}
\label{sec:results}
\HI\ absorption is a powerful tool for providing distance constraints
to Galactic objects \citep[e.g.][]{gathier86,kuchar90,kolpak03}.
Because \HI\ is broadly distributed throughout the Galactic plane, the
presence of \HI\ absorption features (or lack thereof) can constrain
the position of an object behind or in front of specific \HI\
features.  However, interpretation of \HI\ spectra is challenging,
owing to its occasionally patchy distribution
\citep{garwood89}. Furthermore, the application of global circular
rotation curves for estimating distances can be misleading, due
to widespread non-circular gas motions.

In considering \HI\ absorption toward SGR~1806--20, \citet{cameron05}
compared the \HI\ absorption spectrum toward SGR~1806--20 with
 \HI\ and $^{12}$CO emission spectra along the
line of sight, to derive kinematic distance limits to the magnetar.
Figure~2 of \citet{cameron05} presents these absorption and emission
spectra, as well as the \HI\ absorption spectrum toward the nearby
extragalactic source NVSS~J181106--205503  (hereafter J1811--2055).

In general, one uses the absorption feature with the most extreme
velocity (positive or negative) to define a lower limit on the
distance \citep[e.g.][]{kuchar90}.  For Galactic longitudes $0\arcdeg
\leq l \leq 90\arcdeg$, and assuming circular rotation, gas at
negative velocities is more distant from us than gas at positive
velocities.  However, SGR~1806--20 is at a Galactic longitude of
$l=10\arcdeg$ where non-circular motions from the so-called ``3 kpc''
spiral arm produce unusual velocity structures. This can put
relatively nearby emission at negative velocities.  The spectrum of
SGR~1806--20 obtained by \citet{cameron05} is a clear case for which
there is significant \HI\ absorption at a radial velocity $V_{LSR}
\approx - 20$ \kms. As pointed out by \cite{cameron05}, these
features can be attributed to non-circular motions.

The next most extreme velocity feature toward SGR~1806--20 is a
strong absorption line at $V_{LSR}=+85$ \kms.  This agrees well
with both the \HI\ and CO emission at the same velocity.  At
$V_{LSR}=+95$ \kms\ there may be another weak absorption feature,
which \citet{cameron05} state has a significance of $2.5\sigma$.
While we prefer to use the feature at $V_{LSR}=+85$~\kms\ to provide
a definitive lower limit on the distance, we note that the choice
between these two features makes little difference to the inferred
distance, both yielding a lower limit $\sim6$~kpc (see
\S\ref{sec:discussion} below).

A commonly used technique to place an upper limit on \HI\ determined
distances is to compare an absorption spectrum toward the target source
with absorption seen toward nearby extragalactic sources.  If there is
absorption toward an extragalactic source that is from more distant gas
than the last absorption feature observed in the target source, that
can suggest an upper limit for the target source.  The \HI\ absorption
spectrum presented for J1811--2055 by \citet{cameron05} exhibits strong
absorption near $V_{LSR}=+120$ \kms\ which is not present in the spectrum
toward SGR~1806--20.  \citet{cameron05} argue that the absorbing gas seen
at $V_{LSR}=+120$ \kms\ is widespread.  The absence of strong absorption
at $V_{LSR}=+120$ \kms\ toward SGR~1806--20 then provides an upper
limit on the distance of the magnetar.

\begin{figure}[!ht]
\centering
\
\includegraphics[angle=-90,width={\textwidth}]{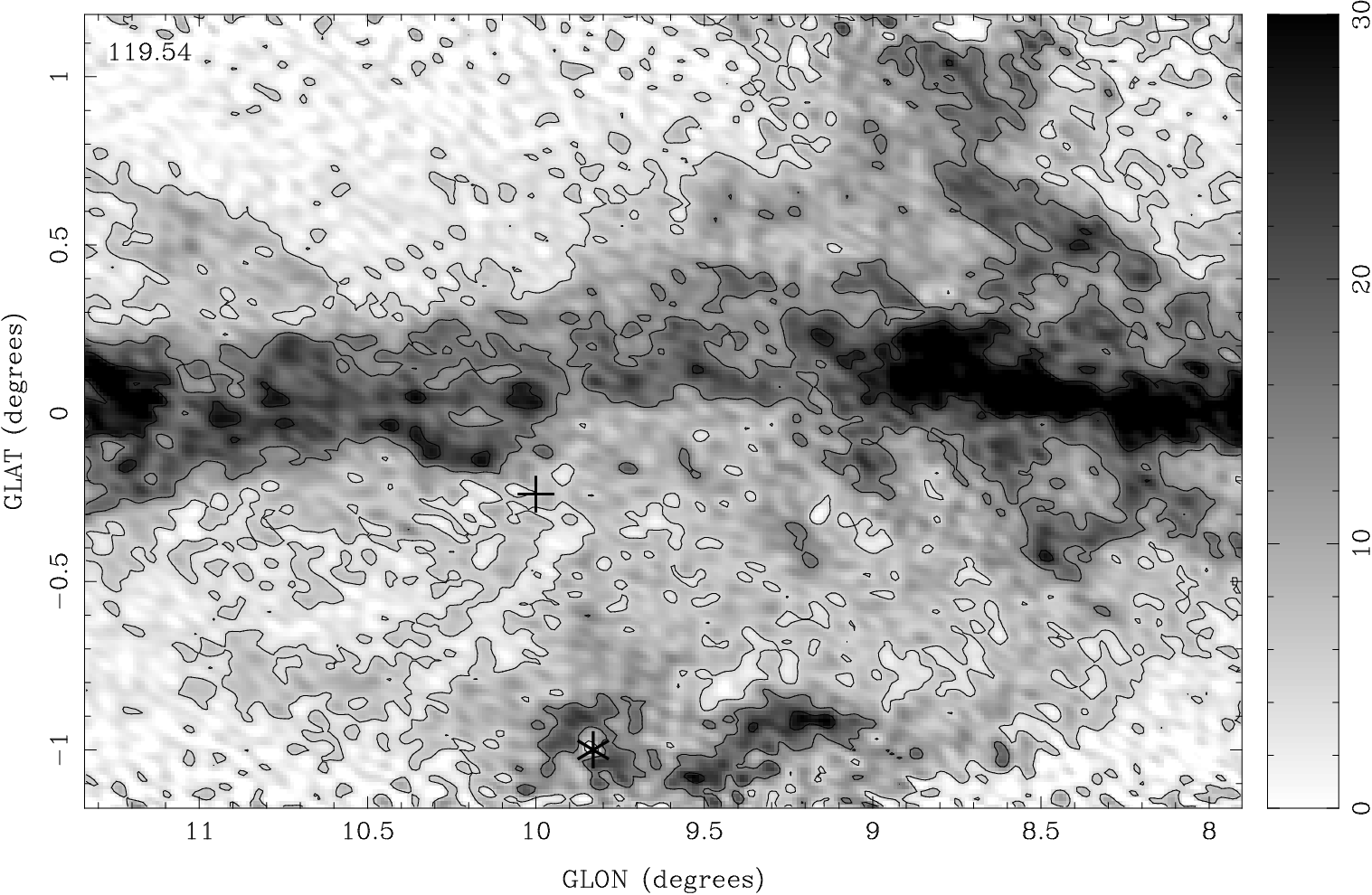}
\caption[]{\HI\ brightness temperature distribution in the region near
  $l=10\arcdeg$, $b=0\arcdeg$ at $V_{LSR}=+120$ \kms\ from the SGPS
  \citep{mcgriff05}.  The image is of a single velocity channel of
  width $\Delta V=0.8$ \kms, and has been continuum subtracted.  The
  greyscale is linear from 0 to 30~K, as shown in the wedge to the
  right. The sensitivity of this image
 is 1.7~K and the spatial resolution is 2~arcmin.  Contour
  levels are at 5~K to 45~K, in intervals of 10 K.  The position of
  SGR~1806--20 is marked with a cross and the position of the
  extragalactic source J1811--2055 is marked with a star.  The \HI\
  distribution is quite inhomogeneous:  there is bright
  emission at the position of J1811--2055, but at the
  position of SGR~1806--20 the \HI\ is almost three times fainter.
  (The apparent depression in the \HI\ distribution at the
  position of J1811--2055 is due to \HI\ absorption against this
  source, and does not represent a real deficit of neutral gas.)
\label{fig:sgps_hi}}
\end{figure}

However, J1811--2055 is nearly one degree away from SGR~1806--20.
The \HI\ distribution can change significantly on scales much smaller
than one degree, and indeed this seems to be the case here.  We
demonstrate this in Figure~\ref{fig:sgps_hi}, where we show an SGPS
image of the \HI\ brightness temperature distribution in the vicinity
of SGR~1806--20 and J1811--2055 at $V_{LSR}=+120$~\kms\ \citep{mcgriff05}.
The positions of SGR~1806--20 and J1811--2055 are marked with a
cross and a star, respectively.  Clearly the \HI\ is not distributed
smoothly  at this velocity.  Of particular importance, the image
shows that there is a strong 
\HI\ feature coincident with J1811--2055
(peak brightness temperature $T_b \sim 30$ K) which
is presumably
also the source of the \HI\ absorption. However, this feature does
not extend to the position of SGR~1806--20, at whose position we
find $T_b \sim 7$~K.\footnote{Conversely, the SGPS \HI\ distribution
at $V_{LSR}=+85$ \kms\ shows \HI\ emission coincident with SGR~1806--20
that is not present at the position of J1811--2055, explaining the
lack of any absorption at $V_{LSR}=+85$~\kms\ toward J1811--2055.}

Additional evidence for this point comes from $^{12}$CO emission,
which is a reasonably good tracer of the cold gas that produces \HI\
absorption.  \citet{garwood89} found that in the inner Galaxy, 85\% of
\HI\ absorption features with optical depths $>0.1$ have counterparts
in $^{12}$CO at the same position and velocity.  A good example of
this agreement is seen in Figure~2 of \citet{cameron05}, where every
\HI\ absorption feature toward SGR~1806--20 is accompanied by a
$^{12}$CO emission feature.  However, the distribution of $^{12}$CO
emission at $V_{LSR}=+120$~\kms\ in this region is patchy, just as is
seen in \HI.  Specifically, the latitude-velocity images of $^{12}$CO
emission in the surveys of \citet{sanders86} and of
\citet{dame01} both clearly show that at $V_{LSR}=+120$~\kms\ there
is $^{12}$CO emission toward J1811--2055, but that there is no CO
emission at this velocity aligned with SGR~1806--20. 
Given the good agreement between \HI\ absorption and
CO emission along these lines of sight at other velocities, it is
difficult to state convincingly that there is cold, absorbing \HI\ at
$V_{LSR}=+120$ \kms\ in the direction of SGR~1806--20.

\section{Conclusion}
\label{sec:discussion}

The absorption feature at $V_{LSR}=+85$ \kms\ seen in the \HI\
spectrum of SGR~1806--20 can be used to place a lower limit on the
distance.  Using the \citet{brand93} rotation curve, with the IAU
standard values $\Theta_0 = 220$ \kms\ and $R_0 = 8.5$~kpc, and
assuming an error in the velocity measurement of $\pm7$~\kms\ to
account for random cloud-to-cloud motions in the ISM
\citep[e.g.,][]{belfort84}, we estimate that SGR~1806--20 must be at a
distance of at least $6.2\pm0.1$~kpc.\footnote{It should be noted that
errors in the assumed rotation curve are likely to significantly
exceed the errors quoted here.  The assumption of circular rotation
breaks down near spiral arms where streaming motions can account for
non-circular motions on the order of $\sim 20$ \kms\ \citep{shane66}.
These non-circular motions are particularly influential in the inner
$\sim 10 - 20$ degrees of the Galaxy, reflecting the effects of the 3
kpc arm and the central bar.  We point this out as a cautionary note
that the minimum kinematic distance could have errors as large
5--7\%.}  However, because \HI\ and $^{12}$CO are both
clearly unevenly distributed at
$V_{LSR}=+120$~\kms, we do not believe that an upper limit to the
distance of SGR~1806--20 can be inferred from the lack of \HI\
absorption at this velocity.
We conclude that the 
distance of SGR~1806--20 as determined from the \HI\ measurements
of \citet{cameron05} is fully consistent with the earlier
distance estimates of $\sim15$~kpc discussed in \S\ref{sec:intro}.  As
considered in detail by \citet{hbs+05} and by \citet{pbg+05}, the
inferred energetics of the giant flare from SGR~1806--20 thus remain
remarkable.

\acknowledgements 
We thank Rob Fender, Greg Taylor, Tom Dame, Dale Frail, Steve Eikenberry,
Shri Kulkarni and Stephane Corbel for useful discussions.  The
Australia Telescope is funded by the Commonwealth of Australia for
operation as a National Facility managed by CSIRO.  B.M.G.  acknowledges
the support of NASA through LTSA grant NAG5-13032.

\small
\bibliographystyle{apj} 
\bibliography{/Users/mcg/tex/references} 
 
\normalsize



\end{document}